\documentclass[11pt,a4paper]{article}
\pdfoutput=1
\usepackage{geometry}
\usepackage{graphicx}
\usepackage{color}
\usepackage{framed}
\usepackage{amssymb}
\usepackage{latexsym}
\usepackage{amsmath}
\usepackage{enumitem}
\usepackage{bm}
\usepackage[width=0.9\textwidth]{caption}
\usepackage[retainorgcmds]{IEEEtrantools}
\usepackage[bookmarks=true,pdfborder={0 0 0}]{hyperref}
\usepackage[all]{hypcap}
\title{Higher charge periodic monopoles}
\author{Rafael Maldonado\footnote{{\tt rafael.maldonado@durham.ac.uk}}\vspace{0.2cm}\\\emph{Department of 
Mathematical Sciences,}\\\emph{South Road, Durham DH1 3LE, UK}}

\renewcommand{\Re}{\text{Re}}
\renewcommand{\Im}{\text{Im}}

\begin{document}
\maketitle
\begin{abstract}
\noindent We consider singly periodic solutions to the SU(2) Bogomolny equations and use the Nahm transform 
to generate a class of monopoles of charge $k>2$, thereby extending known results for lower charge chains.  
Some simple scattering processes are presented and a comparison made with geodesic motion of monopoles in 
$\mathbb{R}^3$.
\end{abstract}
\section{Introduction}\label{introduction}
Solutions of the Bogomolny (monopole) equations on $\mathbb{R}^2\times S^1$ were first considered by Cherkis 
\& Kapustin \cite{ChK01,ChK03}, where the Nahm transform was adapted to this topology.  Approximate solutions 
of charge $1$ and $2$ were then constructed by Harland and Ward \cite{HW09,War05}, and a further class of 
charge $2$ solutions was introduced in \cite{Mal13}.  These charge $2$ examples suggest an Ansatz for higher 
charge solutions, which will be presented in this paper.  If the size to period ratio $C$ is small, the 
results reproduce qualitatively the scattering proccesses described in \cite{HMM95,Sut97}.  We will also 
consider the `spectral approximation' of \cite{Mal13}, which describes the limit of large $C$.
\par The structure of this paper is as follows.  Section \ref{setup} describes the setup specific to this 
case and introduces the spectral curve used in the Nahm construction (the reader is referred to \cite{Mal13} 
for a more general formulation).  In section \ref{solutions} we give a charge $2$ solution to the 
Nahm/Hitchin equations and show how it is generalised to arbitrary charge, considering the symmetries of the 
resulting monopole configurations.  Section \ref{doubledchains} extends a result of \cite{HW09} to show that 
higher charge monopole chains can be constructed from lower charge chains by taking adjacent monopoles in 
pairs.  We wrap up with some conclusions and ideas for further work.
\newpage
\section{Setup}\label{setup}
The periodic monopole can be constructed by use of a generalised Nahm transform, which maps between solutions 
of the $\text{SU}(2)$ Bogomolny equations of charge $k$ on $\mathbb{R}^2\times S^1$, and rank $k$ Hitchin 
data on the dual cylinder $\mathbb{R}\times S^1$.  On the monopole side we define the complex coordinate 
$\zeta=\rho\text{e}^{\text{i}\theta}$ on $\mathbb{R}^2$ and $z\sim z+\beta$ on $S^1$, while on the 
Nahm/Hitchin cylinder we define $r\in\mathbb{R}$ and $t\sim t+2\pi/\beta$ which we combine into a complex 
coordinate $s=r+\text{i}t$.  The Hitchin equations
\begin{equation*}
F_{s\bar{s}}\,=\,-\tfrac{1}{4}[\Phi,\Phi^\dag]\qquad\qquad D_{\bar{s}}\Phi\,=\,\partial_{\bar{s}}\Phi+[A_{\bar{s}},\Phi]\,=\,0\label{hitchin}
\end{equation*}
with $^\dag$ denoting complex conjugate transpose are now to be solved for rank $k$ matrix-valued fields, 
where the characteristic polynomial of $\Phi$ is determined by the spectral curve described in section 
\ref{specc}.  The monopole fields are recovered, up to a gauge, through solutions of the inverse Nahm 
equation,
\begin{equation}
\Delta\Psi\,=\,\begin{pmatrix}\bm{1}_k\otimes(2\partial_{\bar{s}}-z)+2A_{\bar{s}}&\bm{1}_k\otimes\zeta-\Phi\\\bm{1}_k\otimes\bar{\zeta}-\Phi^\dag&\bm{1}_k\otimes(2\partial_s+z)+2A_s\end{pmatrix}\,\Psi\,=\,0\label{invnahm}
\end{equation}
where $\Psi$ is a $(2k\times2)$ matrix $(\Psi_+^\mathsf{T}\,\Psi_-^\mathsf{T})^\mathsf{T}$, subject to the 
normalisation condition
\begin{equation*}
\int_{-\infty}^\infty dr\int_{-\pi/\beta}^{\pi/\beta}dt\,(\Psi^\dag\Psi)\,=\,\bm{1}_2.\label{normalisation}
\end{equation*}
One can then construct the monopole fields using
\begin{equation*}
\hat{\Phi}\,=\,\text{i}\int_{-\infty}^\infty dr\int_{-\pi/\beta}^{\pi/\beta}dt\,(r\Psi^\dag\Psi)\qquad\qquad\hat{A}_i\,=\,\int_{-\infty}^\infty dr\int_{-\pi/\beta}^{\pi/\beta}dt\,(\Psi^\dag\partial_i\Psi).
\end{equation*}
Gauge transformations $\hat{g}$ acting on the monopole fields and $g$ on the Nahm fields transform $\Psi$ as
\begin{equation*}
\Psi(s;\zeta,z)\,\mapsto\,U(s)^{-1}\Psi(s;\zeta,z)\,\hat{g}(\zeta,z),\label{psigauge}
\end{equation*}
a relation we will use to study the spatial symmetries of the monopole corresponding to a given solution of 
the Hitchin equations.  Here, $U=h\otimes g(s)$ and $h$ ~is a constant matrix serving to permute the entries 
of $\Delta$ and those of $\Psi$.
\subsection{Spectral curve}\label{specc}
The spectral curve of a periodic monopole was introduced by Cherkis \& Kapustin \cite{ChK01,ChK03} where an 
equivalence was proven between the `monopole' and `Hitchin' spectral curves, which are polynomials of degree 
$2$ in $w$ and of degree $k$ in $\zeta$.  The monopole spectral curve is the characteristic equation of the 
$z$-holonomy of the monopole fields, $\text{det}(w-V(\zeta))=0$, where $w=\text{e}^{\beta s}$,
\begin{equation}
b_k\zeta^k+b_{k-1}\zeta^{k-1}+\dots+b_1\zeta+(b_0+w+w^{-1})\,=\,0,\label{zspec}
\end{equation}
where the coefficients $b_i$ are independent of $s$.  The Hitchin spectral curve is defined from the Nahm 
data, through $\text{det}(\zeta-\Phi)=0$,
\begin{equation*}
\zeta^k-\zeta^{k-1}\text{tr}(\Phi)+\dots+(-1)^k\text{det}(\Phi)\,=\,0.
\end{equation*}
By matching coefficients of powers of $\zeta$ we obtain the gauge invariants of $\Phi$ as a function of $w$ 
and the moduli encoded by the $\{b_i\}$.  It should be noted that the spectral curve only encodes half of the 
total number of expected moduli (for a centered chain the relative moduli space $\mathcal{M}_k$ has real 
dimension $4k-4$).  In \cite{HW09} and \cite{Mal13} it was shown that in the charge $k=2$ case the moduli 
present in the spectral curve provide a geodesic submanifold of $\mathcal{M}_2$.  These moduli describe the 
relative $xy$ positions of the monopoles but are insensitive to a $z$ offset and relative phase.  For higher 
charges, however, the remaining moduli have yet to be identified.  We appeal to the charge $2$ result and 
assume that the moduli appearing in the spectral curve are the fixed point set of some symmetry group of the 
full moduli space, and thus describe a geodesic submanifold which, following \cite{HMM95}, we denote by 
$\Sigma_k^\ell$, where $\ell$ labels different such submanifolds.
\par It has been found \cite{Mal13} that an approximation to the monopole fields in the limit of large size 
to period ratio (in which the monopole fields are increasingly independent of $z$) can be read off the 
spectral curve polynomial \eqref{zspec} by expressing $s$ in terms of $\zeta$.  We then have 
$\hat{\Phi}=\text{i}\Re(s(\zeta))\sigma_3$, from which the energy density is calculated using
\begin{equation*}
\mathcal{E}\,=\,\partial_\zeta\partial_{\bar{\zeta}}|\text{tr}(\hat{\Phi}^2)|.
\end{equation*}
We will use this result in the following sections to visualise monopole fields with various spatial 
symmetries.
\subsection{Charge 3 symmetries}\label{charge3symmetries}
\begin{figure}
\begin{minipage}{0.485\linewidth}
\centering
\includegraphics[width=1\linewidth]{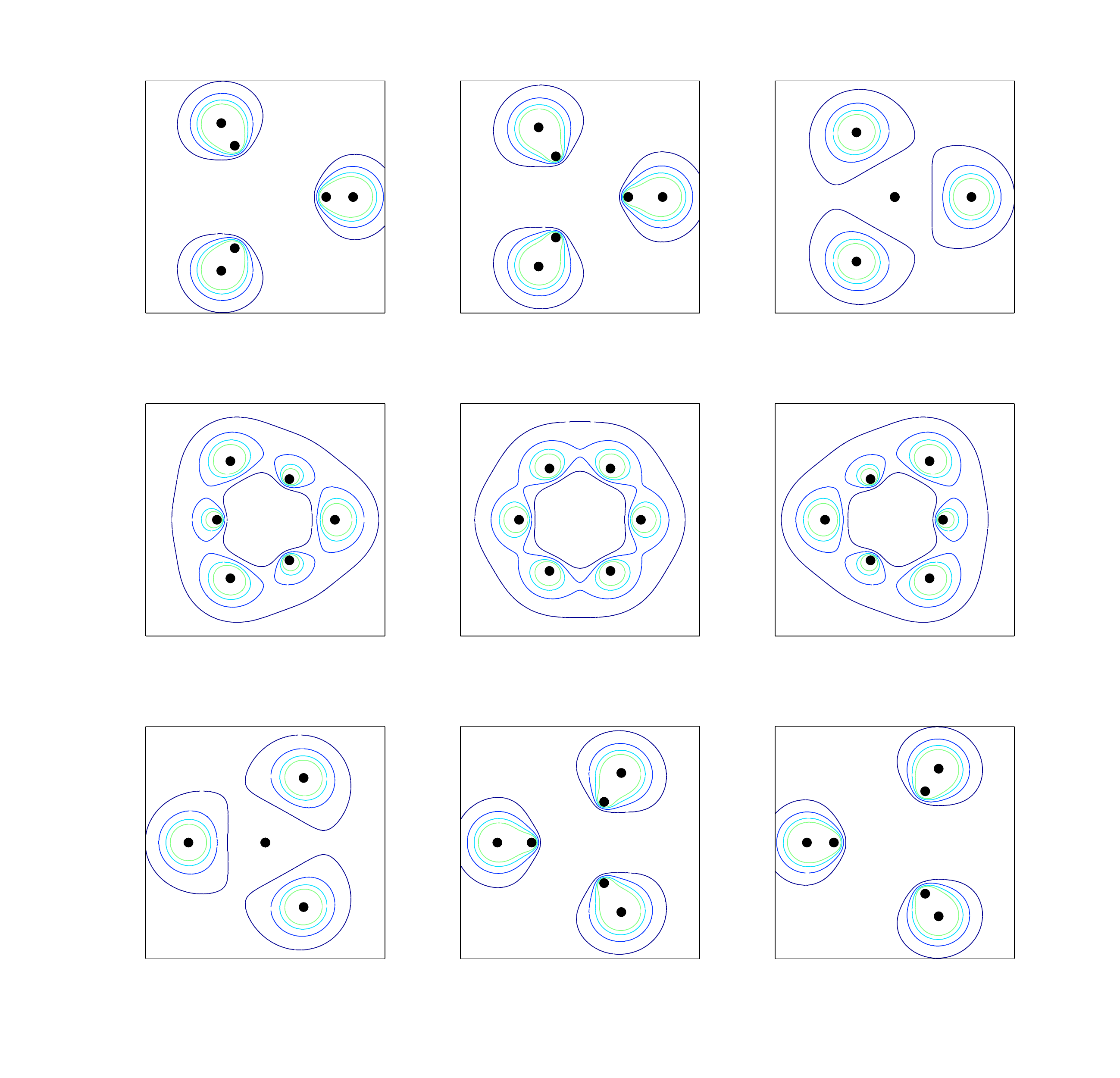}
\end{minipage}
\begin{minipage}{0.485\linewidth}
\centering
\includegraphics[width=1\linewidth]{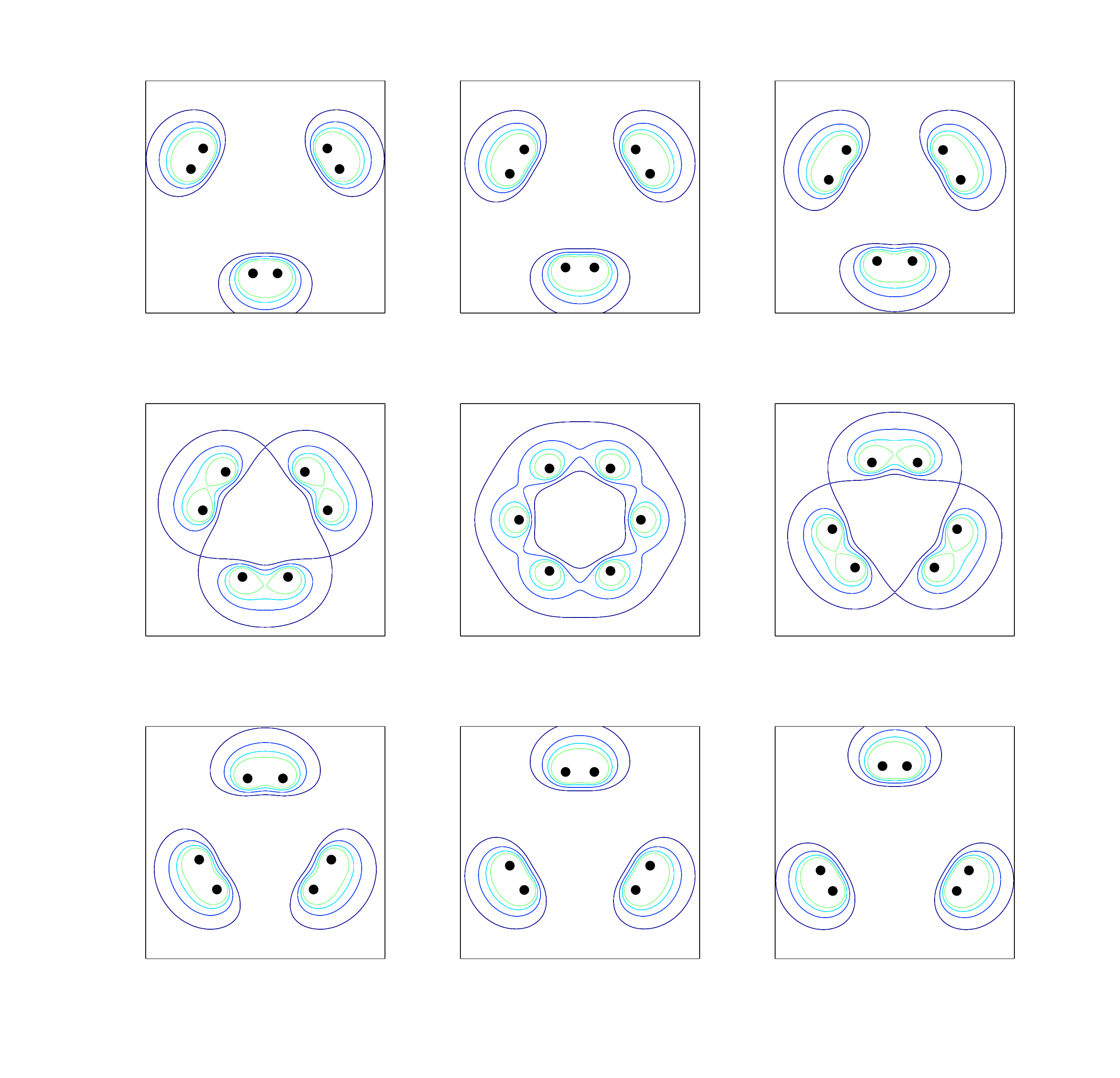}
\end{minipage}
\vspace{-0.5cm}
\caption{Cross section of energy density for two one-parameter families with $b_1=0$.  Left, 
$b_0\in\mathbb{R}$ with $b_0\in[-4,4]$.  The relevant symmetries are \ref{sym1}, \ref{sym2}, \ref{sym4} and 
\ref{sym6} in the list above.  Right, $b_0\in\text{i}\mathbb{R}$ with $-\text{i}b_0\in[-4,4]$, with 
symmetries \ref{sym1}, \ref{sym2}, \ref{sym5} and \ref{sym7}.}\label{geo1}
\end{figure}

\begin{figure}
\centering
\includegraphics[width=0.5\linewidth]{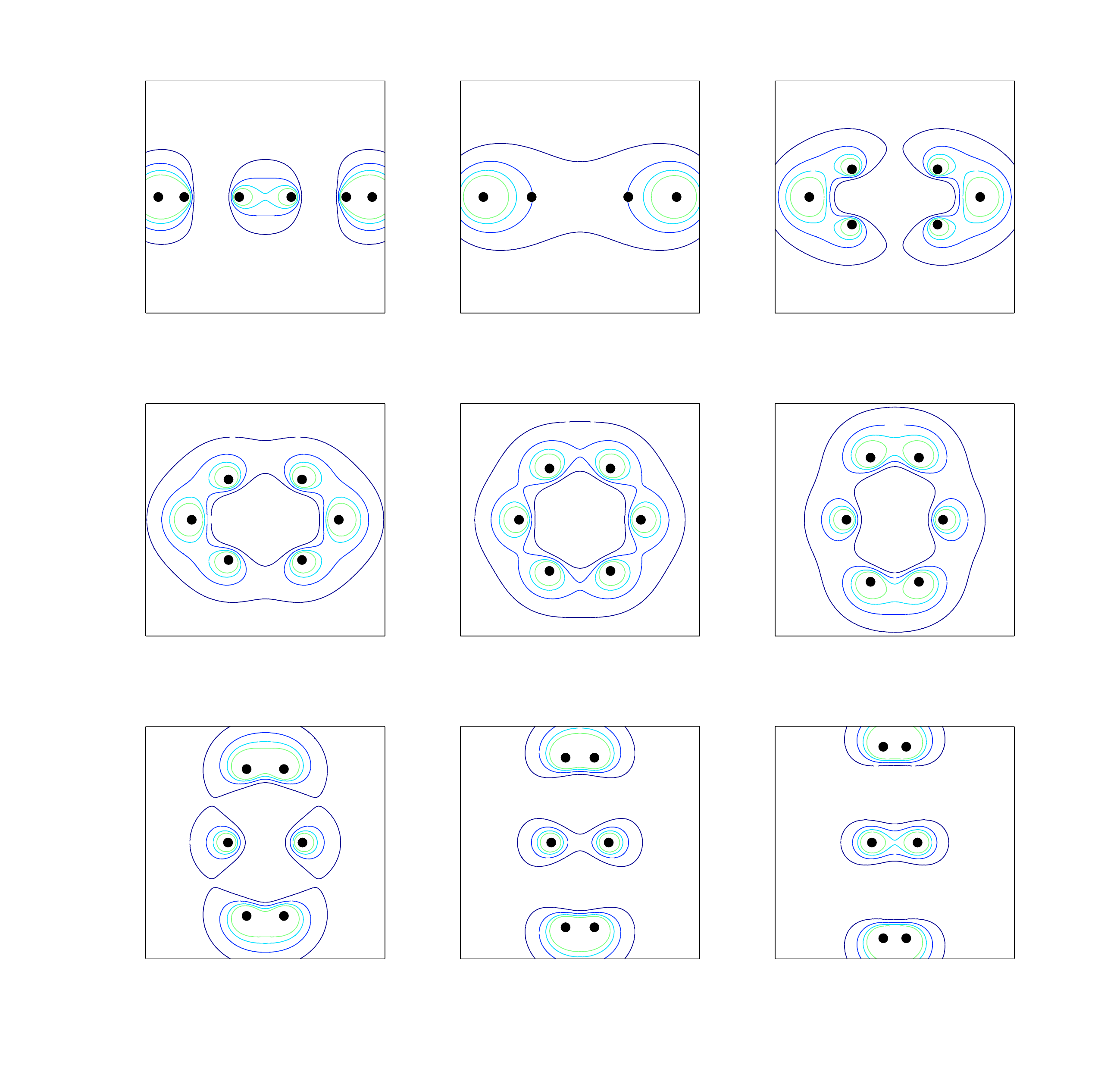}
\vspace{-0.5cm}
\caption{Energy density for $b_1\in\mathbb{R}$, $b_0=0$, with $b_1\in[-4,4]$.  Unlike the symmetries in 
fig.~\ref{geo1}, this family does not have a charge $2$ analogy, and in fact the Nahm data is only known for 
the special case $b_1=-3$, $b_0=0$ (top middle picture).  This configuration is in fact a charge $1$ chain 
with period $\beta/3$ and is described in section \ref{rescaling}.  The relevant symmetries are \ref{sym1}, 
\ref{sym3}, \ref{sym4} and \ref{sym7}.}\label{geo3}
\end{figure}

Geodesic submanifolds can be identified by considering the fixed point sets of symmetries of the spectral 
curve.  We consider two transformations of $\zeta$ (corresponding to a rotation by $\alpha$ and a reflection 
in the line $\theta=\alpha/2$), and find the necessary maps of the coefficients $b_i$ such that the original 
spectral curve is recovered.  The $k=3$ spectral curve can be written
\begin{equation}
w^2+w(b_3\zeta^3+b_2\zeta^2+b_1\zeta+b_0)+1\,=\,0.\label{ch3spec}
\end{equation}
We take $b_3=1$ for the rest of this section, its magnitude setting a scale and its phase an orientation.  We 
also fix the centre of mass of the spectral points at the origin by setting $b_2=0$.\\ \\
$\underline{\zeta\mapsto\zeta\text{e}^{\text{i}\alpha}}$\\ \\
To keep the spectral curve invariant we transform $w\mapsto w\text{e}^{-3\text{i}\alpha}$ and look for values 
of $\alpha$ for which the resulting spectral curve,
\begin{equation*}
w^2\text{e}^{-6\text{i}\alpha}+w\left(\zeta^3+b_1\zeta\text{e}^{-2\text{i}\alpha}+b_0\text{e}^{-3\text{i}\alpha}\right)+1\,=\,0,
\end{equation*}
is the same as the original one, \eqref{ch3spec}, for a certain choice of $b_1$ and $b_0$.
\begin{enumerate}[label=\roman{*}., ref=\roman{*}]
\item $\alpha=\pi/3$, $b_1\mapsto\text{e}^{2\text{i}\pi/3}b_1$, $b_0\mapsto-b_0$, with fixed set $b_1=b_0=0$.  
This corresponds to the hexagonally symmetric configuration of spectral points.\label{sym1}\vspace{-0.2cm}
\item $\alpha=2\pi/3$, $b_1\mapsto\text{e}^{4\text{i}\pi/3}b_1$, $b_0\mapsto b_0$, with fixed set $b_1=0$ for 
all $b_0$.\label{sym2}\vspace{-0.2cm}
\item $\alpha=\pi$, $b_1\mapsto b_1$, $b_0\mapsto-b_0$, with fixed set $b_0=0$ for all $b_1$.\label{sym3}
\end{enumerate}
$\underline{\zeta\mapsto\bar{\zeta}\text{e}^{\text{i}\alpha}}$\\ \\
We also set $w\mapsto\bar{w}\text{e}^{-3\text{i}\alpha}$, such that
\begin{equation*}
\bar{w}^2\text{e}^{-6\text{i}\alpha}+\bar{w}\left(\bar{\zeta}^3+b_1\bar{\zeta}\text{e}^{-2\text{i}\alpha}+b_0\text{e}^{-3\text{i}\alpha}\right)+1\,=\,0
\end{equation*}
\begin{equation*}
w^2\text{e}^{6\text{i}\alpha}+w\left(\zeta^3+\bar{b}_1\zeta\text{e}^{2\text{i}\alpha}+\bar{b}_0\text{e}^{3\text{i}\alpha}\right)+1\,=\,0.
\end{equation*}
\begin{enumerate}[label=\roman{*}., ref=\roman{*}]
\setcounter{enumi}{3}
\item $\alpha=0$, $b_1\mapsto\bar{b}_1$, $b_0\mapsto\bar{b}_0$, with fixed set $b_1\in\mathbb{R}$ and 
$b_0\in\mathbb{R}$.\label{sym4}\vspace{-0.2cm}
\item $\alpha=\pi/3$, $b_1\mapsto\text{e}^{2\text{i}\pi/3}\bar{b}_1$, $b_0\mapsto-\bar{b}_0$, with fixed set 
$b_1=\text{e}^{\text{i}\pi/3}|b_1|$, $b_0\in\text{i}\mathbb{R}$.\label{sym5}\vspace{-0.2cm}
\item $\alpha=2\pi/3$, $b_1\mapsto\text{e}^{-2\text{i}\pi/3}\bar{b}_1$, $b_0\mapsto\bar{b}_0$, with fixed set 
$b_1=\text{e}^{-\text{i}\pi/3}|b_1|$, $b_0\in\mathbb{R}$.\label{sym6}\vspace{-0.2cm}
\item $\alpha=\pi$, $b_1\mapsto\bar{b}_1$, $b_0\mapsto-\bar{b}_0$, with fixed set $b_1\in\mathbb{R}$ and 
$b_0\in\text{i}\mathbb{R}$.\label{sym7}
\end{enumerate}
The above symmetries of the spectral curve can be combined to give three distinct scattering processes, 
described in figs \ref{geo1} and \ref{geo3}.
\newpage
\section{Constructing solutions}\label{solutions}
We start this section by recalling the charge $2$ solutions to the Nahm equations described in 
\cite{HW09,Mal13}, and then give an extension to higher charges, making specific reference to the charge $3$ 
and $4$ cases as an illustration.
\subsection{Charge 2}\label{charge2}
For $k=2$ the spectral curve polynomials require that the Hitchin Higgs field $\Phi$ is of rank $2$ and 
satisfies
\begin{equation*}
\text{tr}(\Phi)\,=\,0\qquad\qquad-\text{det}(\Phi)\,=\,C\cosh(\beta s)-K/2
\end{equation*}
where $C$ and $K$ are defined in terms of the coefficients in \eqref{zspec} by $C=-2/b_2$ and $K=2b_0/b_2$, 
such that $C$ is fixed by the boundary conditions and describes the size and orientation of the monopole (we 
will take it to be real and positive) and $K$ is a complex modulus encoding the positions of the monopoles in 
the $xy$ plane.  In \cite{HW09} it was shown that the general solution to the Hitchin equations is (up to a 
gauge)
\begin{equation*}
\Phi\,=\,\begin{pmatrix}0&\mu_+\text{e}^{\psi/2}\\\mu_-\text{e}^{-\psi/2}&0\end{pmatrix}\qquad\qquad A_{\bar{s}}\,=\,a\sigma_3+\alpha\Phi\qquad\qquad A_s\,=\,-A_{\bar{s}}^\dag,
\end{equation*}
where $4a=-\partial_{\bar{s}}\psi$,
\begin{equation*}
\nabla^2\Re(\psi)\,=\,2(1+4|\alpha|^2)\left(|\mu_+|^2\text{e}^{\Re(\psi)}-|\mu_-|^2\text{e}^{-\Re(\psi)}\right)
\end{equation*}
and
\begin{equation*}
\text{e}^{-\Re(\psi)/2}\partial_s\left(\alpha\mu_+\text{e}^{\Re(\psi)}\right)+\text{e}^{\Re(\psi)/2}\partial_{\bar{s}}\left(\bar{\alpha}\bar{\mu}_-\text{e}^{-\Re(\psi)}\right)\,=\,0.
\end{equation*}
Imposing that the monopole fields have the spatial symmetry $(\zeta,z)\sim(-\zeta,z)$ fixes $\alpha=0$, as 
can be seen by gauge transforming the Hitchin fields by $\sigma_3$ along with $\Psi_\pm\mapsto\pm\Psi_\pm$ in 
the inverse Nahm operator \eqref{invnahm}.  The function $\alpha$ is expected to encode the two remaining 
moduli: a $z$ offset and relative phase.
\par The key point is that $C\cosh(\beta s)-K/2$ has two zeroes, whose positions depend on the value of 
$K/C$.  There are then two classes of smooth solutions of the Hitchin equations, according to the allocation 
of zeroes between $\mu_+$ and $\mu_-$, such that $\ell=0$ if both zeroes are in $\mu_+$ and $\ell=1$ if one 
is in each of $\mu_\pm$ (see section \ref{specc}).  Geodesics on each of these submanifolds are studied by 
transforming $s$ and $K$ in such a way that the transformed fields can be written as a gauge transformation 
of the original fields and then observing the effect of this change on the monopole coordinates $\zeta$ and 
$z$ via \eqref{invnahm} (for more details, see \cite{Mal13}).
\begin{itemize}
\item The $\ell=0$ solution has
\begin{equation}
\mu_+\,=\,C\cosh(\beta s)-K/2\qquad\qquad\mu_-\,=\,1\label{zeroestogether}
\end{equation}
with $\Im(\psi)=0$.  The geodesics $K\in\mathbb{R}$ and $K\in\text{i}\mathbb{R}$ describe $\pi/2$ scattering 
of monopoles in the $xy$ plane, via a toroidal central configuration with discrete $\mathbb{Z}_4$ symmetry.
\item The $\ell=1$ configuration has
\begin{equation}
\mu_\pm\,=\,\sqrt{C/2}\left(\text{e}^{\beta s/2}-W^{\pm1}\text{e}^{-\beta s/2}\right)\qquad\text{with}\qquad K/C\,=\,W+W^{-1}\label{zeroesapart}
\end{equation}
and $\Im(\psi)=-\beta t$ in order for $\Phi$ to be periodic with $t\sim t+2\pi/\beta$.  Simple geodesics 
representing monopole scattering are obtained as fixed points of the symmetries of the Hitchin equations, as 
described in \cite{MW} (note that \cite{HW09,Mal13} interpret the branch structure differently).  Two of them 
correspond to double scattering, first along $z$ and then in the $xy$ plane (either parallel or at right 
angles to the incoming monopoles), while $|W|=1$ is a closed geodesic with monopoles fixed at $z=\pm\beta/4$ 
but oscillating in shape.
\end{itemize}
\subsection{Higher charges}\label{highercharges}
A straightforward extension of the charge $2$ solutions described in the previous section is a modification 
of ``Sutcliffe's ansatz'' \cite{Sut96,Bra11}.  Solutions generated in this way have $b_i=0$ for $i\neq0,k$ in 
\eqref{zspec}.  We take
%
\begin{equation}
\Phi\,=\,\begin{pmatrix}0&0&\cdots&0&f_1\\f_2&0&\cdots&0&0\\0&f_3&\cdots&0&0\\\vdots&\vdots&\ddots&\vdots&\vdots\\0&0&\cdots&f_k&0\end{pmatrix}\qquad\qquad A_{\bar{s}}\,=\,\begin{pmatrix}a_1&0&0&\dots&0\\0&a_2&0&\dots&0\\0&0&a_3&\dots&0\\\vdots&\vdots&\vdots&\ddots&\vdots\\0&0&0&\dots&a_k\end{pmatrix}.\label{Phin}
\end{equation}
Mimicking the charge $2$ procedure, we define $f_i=\mu_i\text{e}^{\psi_i/2}$, with the conditions 
$\sum_{i=1}^k\psi_i=0$ and $\prod_{i=1}^k\mu_i=(-1)^{k-1}\text{det}(\Phi)$.  The Hitchin equations then become
\begin{equation*}
2\left(a_{i-1}-a_i\right)\,=\,\partial_{\bar{s}}\psi_i
\end{equation*}
\begin{equation*}
\nabla^2\log|f_i|^2\,=\,2|f_i|^2-|f_{i-1}|^2-|f_{i+1}|^2\label{fequation}
\end{equation*}
where the index $i$ is periodic, such that $f_0=f_k$.  As was the case in section \ref{charge2}, the 
determinant of $\Phi$ has exactly two zeroes, such that smooth solutions must have both zeroes in the same or 
different entries $\mu_i$ (such that two of the $\mu_i$ are $\mu_\pm$ and all the others are set to $1$).  We 
are free to fix one of the zeroes, $\mu_1=\mu_+$.  Then for given charge $k$, the $\ell=0$ configuration 
has both zeroes in $\mu_1$, and there are $(2k+(-1)^k-1)/4$ gauge inequivalent configurations with 
$\ell>0$, where $\ell$ is the separation between the positions of $\mu_\pm$ in $\Phi$, so in particular 
$\mu_{1+\ell}=\mu_-$.  
With this notation, the Hitchin equations for $k=3$, $\ell=1$ are
\begin{equation}
\left\lbrace\begin{array}{rcl}\nabla^2\Re(\psi_1)\,&=&\,2|\mu_+|^2\text{e}^{\Re(\psi_1)}-|\mu_-|^2\text{e}^{\Re(\psi_2)}-\text{e}^{-\Re(\psi_1+\psi_2)}\\\nabla^2\Re(\psi_2)\,&=&\,2|\mu_-|^2\text{e}^{\Re(\psi_2)}-|\mu_+|^2\text{e}^{\Re(\psi_1)}-\text{e}^{-\Re(\psi_1+\psi_2)}\end{array}\right.\label{ch3psieqs}
\end{equation}
with $\mu_\pm$ as in \eqref{zeroestogether} or \eqref{zeroesapart}.
\par Solving the Hitchin equations numerically is now a matter of adapting the charge $2$ procedure used in 
\cite{HW09}.  First of all we note the equations \eqref{ch3psieqs} can be obtained by varying the functional
\begin{equation}
E[\Re(\psi_i)]\,=\,\int dr\,dt\left(\frac{1}{2}\sum_{p=r,t}\left(\partial_p\Re(\psi_i)\right)^2+2|\mu_i|^2\text{e}^{\Re(\psi_i)}-\psi_i|\mu_j|^2\text{e}^{\Re(\psi_j)}+\text{e}^{-\Re(\psi_i+\psi_j)}\right)\label{psifunctional}
\end{equation}
with respect to $\psi_i$, where $i,j\in\{1,2|i\neq j\}$ and no sum is implied.  Unfortunately there appears 
to be no simple way of combining the two functionals which generate the separate equations \eqref{ch3psieqs} 
into a single expression.  Instead of minimising a single functional, we alternately minimise 
$E[\Re(\psi_1)]$ and $E[\Re(\psi_2)]$.  This approach was found to lead to rapidly convergent solutions as 
long as the boundary conditions were chosen appropriately.  In fact, it is straightforward to write down an 
asymptotic solution to \eqref{ch3psieqs} valid away from the zeroes of $\mu_\pm$ by making the Ansatz 
$\psi_i=\log(|\mu_+|^{\nu^+_i}|\mu_-|^{\nu^-_i})$ and solving for the $\nu^\pm_i$ (this solution is singular 
at the zeroes of $\mu_\pm$ and is thus not globally valid).  For $k=3$ and $\ell=1$ we find
\begin{equation*}
\Re(\psi_1)\,=\,\frac{2}{3}\log\frac{|\mu_-|}{|\mu_+|^2}\qquad\Re(\psi_2)\,=\,\frac{2}{3}\log\frac{|\mu_+|}{|\mu_-|^2}\qquad\Re(\psi_3)\,=\,\frac{2}{3}\log\left(|\mu_+||\mu_-|\right).
\end{equation*}
\par There is some freedom in the choice of imaginary parts of the functions $\psi_i$, which must be chosen 
so as to make the Nahm data periodic on the cylinder.  We fix $\Im(\psi_1)=-\beta t$, 
$\Im(\psi_{1+\ell})=\beta t$ and $\Im(\psi_3)=0$.  A different choice simply corresponds to a global shift in 
the $z$ direction, and the resulting moduli space is isomorphic to $\Sigma_3^1$.
\par One might also be concerned by the fact that \eqref{psifunctional} is not explicitly positive definite 
due to the term linear in $\psi_i$, which does not appear in the charge $2$ case.  We again resort to the 
convergence of the numerical solution to justify this approach.
\par It is easy to see that there are no solutions with $\Re(\psi_1)=\Re(\psi_2)=0$ everywhere.  This tells 
us that the charge $1$ chain of period $\beta/3$ is not included in this family of solutions, as this 
would require $F=0$ (see also section \ref{rescaling}).
\subsection{Symmetries}\label{symmetries}
Spatial symmetries of the monopole fields can be studied by the procedure outlined in section \ref{charge2}.  
First of all we choose a transformation of $K$ (or $W$) and $s$ which preserves the spectral curve for a 
given transformation of $\zeta$.  Then we express the transformed Hitchin fields as a gauge transformation of 
the original fields.  This allows us to read off the corresponding change in $z$ from the inverse Nahm 
operator \eqref{invnahm}.
\par Note that if we restrict to gauge transformations which change the positions and phases of the entries 
of $\Phi$, then the overall ordering of the $f_i$ is unchanged (or reversed in the case of $\Phi^\dag$).  
This property gives the solutions $\ell=0$ and $\ell=k/2$ (for $k$ even) an additional $s\mapsto-s$ 
symmetry (corresponding to $z\mapsto-z$), which is not observed for general $\ell$.
\par Various scattering processes generalising those in section \ref{charge2} are described in the following 
subsections, and we visualise them with reference to chains of small monopoles ($C\lesssim 1$).  In summary, 
it is found that the geodesics are characterised by the positions of the zeroes among the entries of $\Phi$, 
say at $f_1$ and $f_{1+\ell}$.  Then for $|W|>1$ the monopoles are located on the vertices of a regular 
$k$-gon at $z=\beta\ell/k$ (the $z$ position is determined numerically).  As $|W|$ is reduced they scatter 
and split into two clusters of charge $\ell$ moving along the positive $z$ axis and $(k-\ell)$ along the 
negative $z$ axis.  The clusters move at speeds inversely proportional to their charges, such that for 
$|W|<1$ the outgoing monopoles emerge at $z=0$ on a (possibly rotated) $k$-gon.  Following the discussion of 
\cite{MW} we expect there to be a closed geodesic with $|W|=1$, describing stationary monopoles oscillating 
in shape.  A discussion of the motion of Higgs zeroes is given in section \ref{Higgszeroes}.
\subsubsection{Planar scattering}\label{planar}
The geodesic surface $\Sigma_k^0$ with $K\in\mathbb{R}$ or $K\in\text{i}\mathbb{R}$ describes scattering in 
the $xy$ plane via a $\mathbb{Z}_{2k}$-symmetric toroidal configuration.  We see this as follows:
\par First of all, note that under the transformation $s\mapsto-s$, $\mu_\pm$ and $\psi$ are invariant and 
$a_i(s)\mapsto a_i(-s)=-a_i(s)$.  The form of the inverse Nahm operator \eqref{invnahm} now tells us that the 
monopole fields are invariant if we also replace $z$ by $-z$.  Thus, this monopole configuration has the 
symmetry $(\zeta,z)\sim(\zeta,-z)$, consistent with the $k$ incoming monopoles being located at $z=0$ (the 
fixed point set of this transformation).
\par To see the $\mathbb{Z}_{2k}$ symmetry we perform the transformation 
$(s;K)\mapsto(s+\text{i}\frac{\pi}{\beta};-K)$, giving $\mu_\pm\mapsto\mp\mu_\pm$ and $\psi_i\mapsto\psi_i$.  
Then $\Phi'(s,K)=\Phi(s+\text{i}\frac{\pi}{\beta},-K)$ is the same as $\Phi(s;K)$ but with the sign of $f_1$ 
reversed.  Under a suitably chosen diagonal gauge transformation $g$, we then have 
$\Phi'=\text{e}^{\text{i}\pi/k}g^{-1}\Phi g$, leaving $A$ unchanged.  The entry $(\zeta-\Phi)$ in the inverse 
Nahm operator \eqref{invnahm} implies that $\zeta\mapsto\zeta\text{e}^{\text{i}\pi/k}$ when we map $K$ to 
$-K$.  The monopole fields are symmetric under $(\zeta,z;K)\mapsto(\zeta\text{e}^{\text{i}\pi/k},z;-K)$, and 
thus $K=0$ describes a configuration of enhanced symmetry.
\subsubsection{Symmetric splitting}\label{symmetricsplitting}
For even $k$, the geodesic submanifold $\Sigma_k^{k/2}$ describes a splitting of $k$ incoming monopoles into 
two equal clusters.  In the case of $\Sigma_4^2$ we identify the following symmetries:
\begin{itemize}
\item $(s,W)\mapsto(-\bar{s},\overline{W})$ $\Rightarrow$ $(\zeta,z)\mapsto(\bar{\zeta},z)$, fixing the 
geodesic $W\in\mathbb{R}$.
\item $(s,W)\mapsto(\text{i}\pi/\beta-\bar{s},-\overline{W})$ $\Rightarrow$ 
$(\zeta,z)\sim(\text{e}^{\text{i}\pi/4}\bar{\zeta},z)$, with fixed point set $W\in\text{i}\mathbb{R}$.
\item $(s,W)\mapsto(\bar{s},\overline{W}^{-1})$ $\Rightarrow$ $(\zeta,z)\mapsto(\bar{\zeta},\beta/2-z)$, 
relates the incoming and outgoing legs of the geodesics $W\in\mathbb{R}$ and $W\in\text{i}\mathbb{R}$.  Thus, 
$W\in\mathbb{R}$ describes monopoles incoming and outgoing parallel to the $x$ and $y$ axes, with a 
half-period shift along $z$.  On the other hand, $W\in\text{i}\mathbb{R}$ has an additional $\pi/4$ rotation 
about the $z$ axis.  This symmetry also fixes the closed geodesic $|W|=1$.
\item $(\zeta,z)\mapsto(\text{i}\zeta,z)$ is a symmetry for all $W$, as can be seen by the gauge 
transformation $g=\text{diag}(1,\text{i},-1,-\text{i})$.
\item $s\mapsto-s$ $\Rightarrow$ $(\zeta,z)\mapsto(\zeta,-z)$ for all $W$.
\end{itemize}
There are two particularly symmetric cases which will be considered in more detail in section 
\ref{doubledchains},
\begin{itemize}
\item $W=1$ has $(\zeta,z)\sim(\zeta,\beta/2-z)\sim(\zeta,z+\beta/2)$,
\item $W=\text{i}$ has $(\zeta,z)\sim(\text{e}^{\text{i}\pi/4}\zeta,\beta/2-z)\sim(\text{e}^{\text{i}\pi/4}\zeta,z+\beta/2)$.
\end{itemize}
The fixed points of these symmetries tell us that the clusters are located at $z=\pm\beta/4$.
\subsubsection{Generic $\ell$}\label{generic}
Here we consider the example of $\Sigma_3^1$.  The symmetries are
\begin{itemize}
\item $(s,W)\mapsto(-\bar{s},\overline{W})$ $\Rightarrow$ $(\zeta,z)\mapsto(\bar{\zeta},z)$, for $W\in\mathbb{R}$.
\item $(s,W)\mapsto(\text{i}\pi/\beta-\bar{s},-\overline{W})$ $\Rightarrow$ $(\zeta,z)\sim(-\bar{\zeta},z)$, for $W\in\text{i}\mathbb{R}$.
\item $(s,W)\mapsto(\bar{s},\overline{W}^{-1})$ $\Rightarrow$ $(\zeta,z)\mapsto(\bar{\zeta},\beta/3-z)$.
\item $(\zeta,z)\mapsto(\text{e}^{2\text{i}\pi/3}\zeta,z)$ is a symmetry for all $W$.
\end{itemize}
In this case, there is no symmetry $z\mapsto-z$ due to the asymmetric splitting.  There are still points with 
enhanced symmetry,
\begin{itemize}
\item $W=1$ has $(\zeta,z)\sim(\zeta,\beta/3-z)$,
\item $W=\text{i}$ has $(\zeta,z)\sim(-\zeta,\beta/3-z)$,
\end{itemize}
with fixed points at $z=\beta/6$ and $2\beta/3$, which are the positions of the charge $2$ and charge $1$ 
clusters.
\par
\begin{figure}
\begin{minipage}{0.485\linewidth}
\centering
\includegraphics[width=1\linewidth]{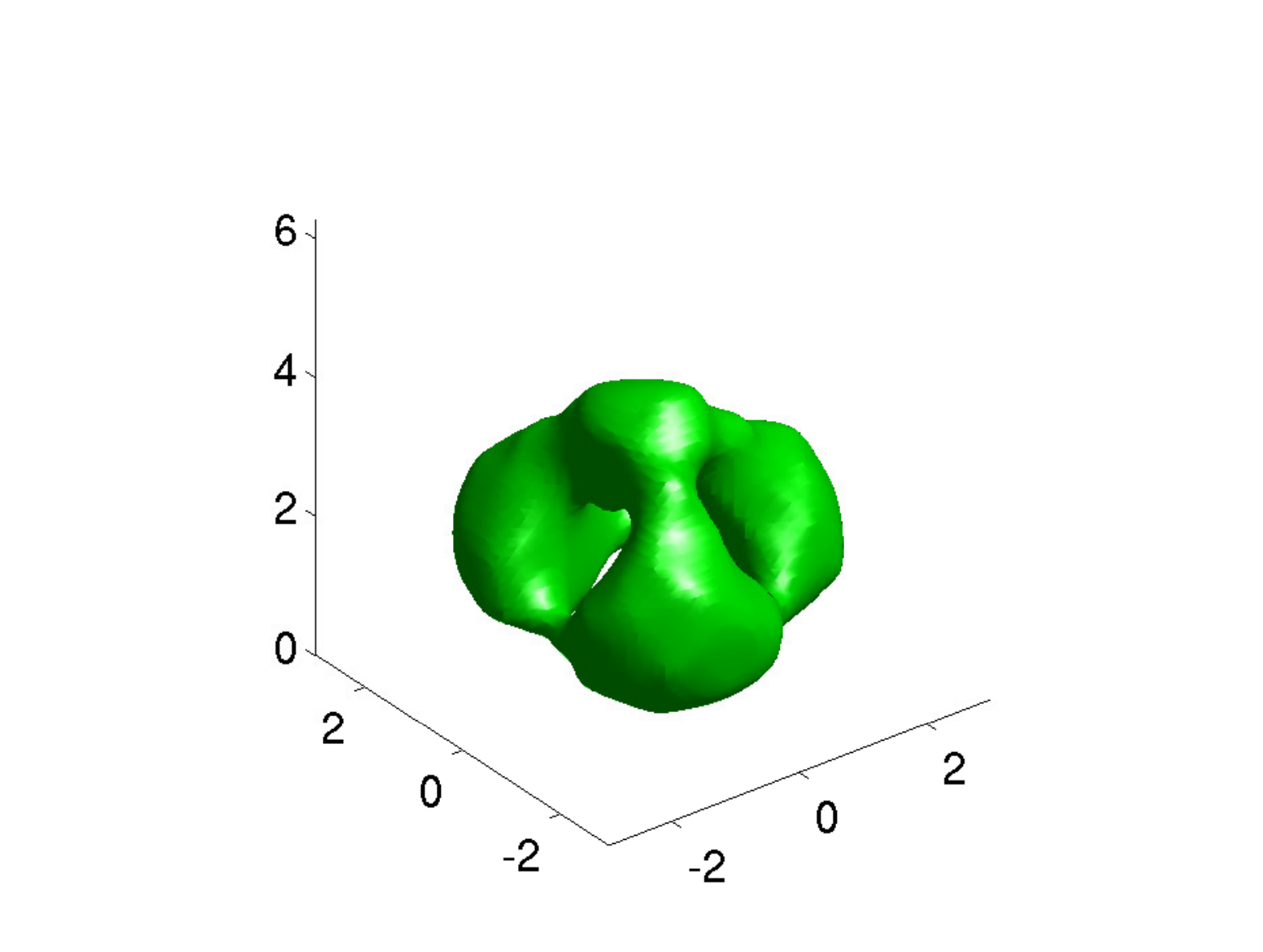}
\end{minipage}
\begin{minipage}{0.485\linewidth}
\centering
\includegraphics[width=1\linewidth]{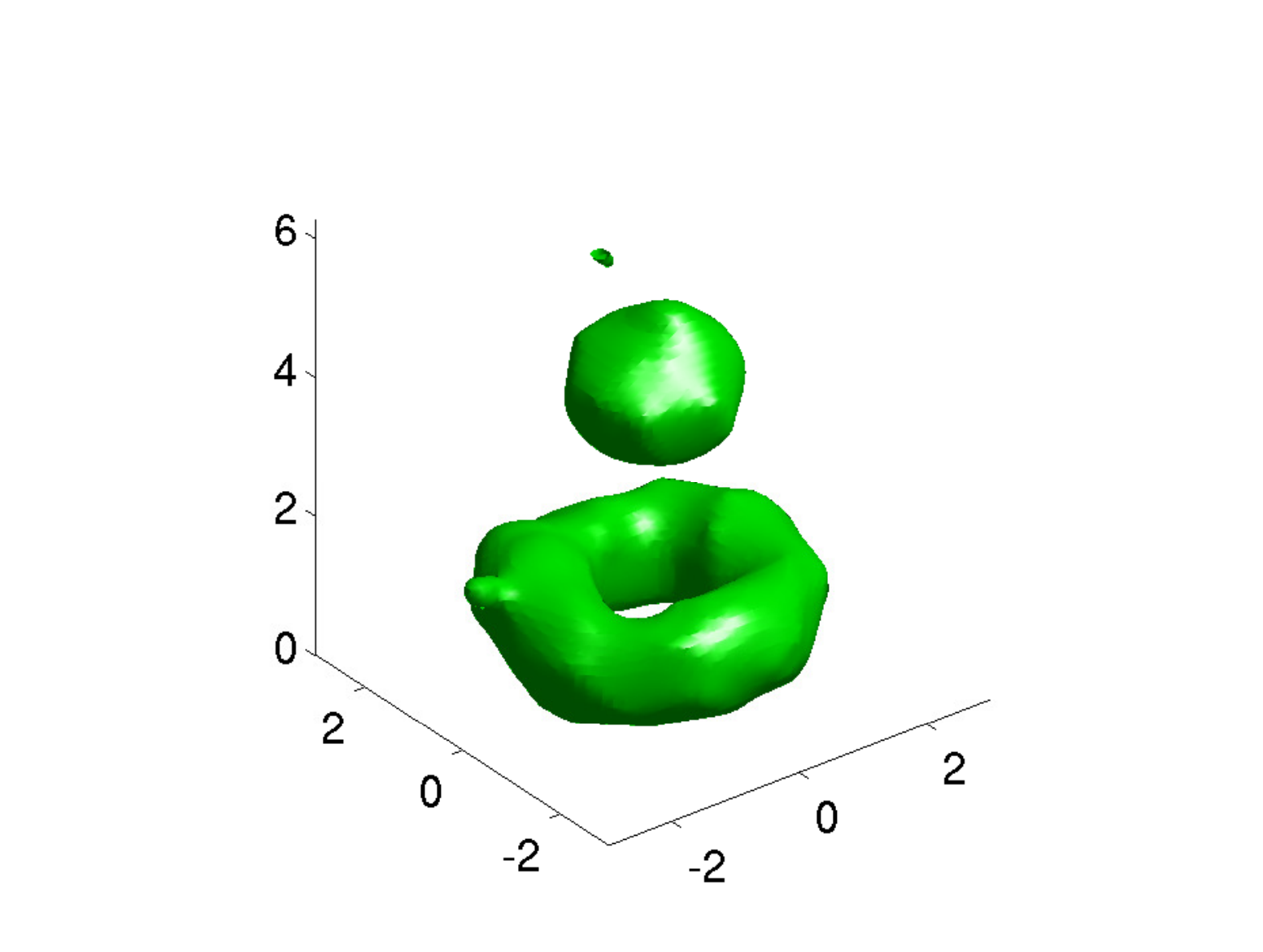}
\end{minipage}
\vspace{-0.2cm}
\caption{Energy density of a charge $3$ periodic monopole with $C=1$ taken over a single period.  Left: 
approximately tetrahedral configuration with $W=2+\sqrt{3}$ ($K=4$).  Right: clusters of charge $1$ and $2$ 
are visible at $W=\text{i}$.}\label{charge3energy}
\end{figure}

 These symmetries are consistent with the expected scattering process.  Monopoles are incoming on the 
vertices of an equilateral triangle.  They scatter along $z$ via an approximately tetrahedral configuration 
to form two clusters (fig.~\ref{charge3energy}).  A new tetrahedral configuration forms from clusters in 
adjacent periods, and outgoing monopoles are shifted by $\beta/3$ and are either rotated by $\pi/3$ about the 
$z$-axis (for $W\in\text{i}\mathbb{R}$) or move back along the original directions (for $W\in\mathbb{R}$).
\subsection{Higgs zeroes}\label{Higgszeroes}
\begin{figure}
\centering
\includegraphics[width=0.7\linewidth]{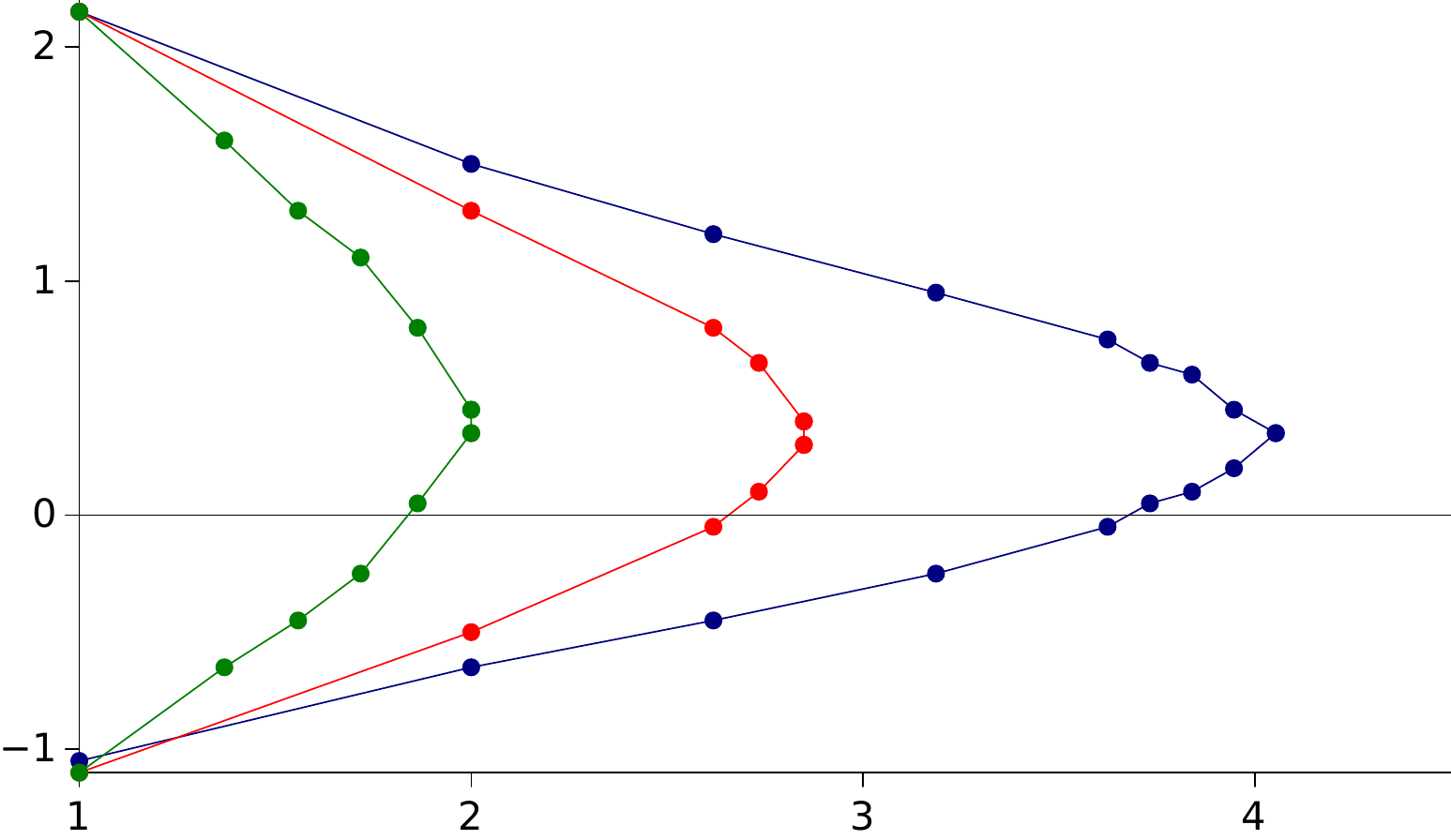}
\caption{Motion of the zero-antizero pair along the $z$-axis (with $0$ corresponding to $z=\beta/3$ and 
$\beta=2\pi$) as a function of $W$ for various values of $C$: $C=1$ in blue (rightmost curve), $C=2$ in red 
(middle) and $C=5$ in green (left).  For small $C$, the value of $W$ at which the lower zero (the anitzero) 
is centered at $z=\beta/3$ roughly coincides with the monopole closest to tetrahedral 
symmetry.}\label{zeroantizero}
\end{figure}

\begin{figure}
\centering
\includegraphics[width=0.5\linewidth]{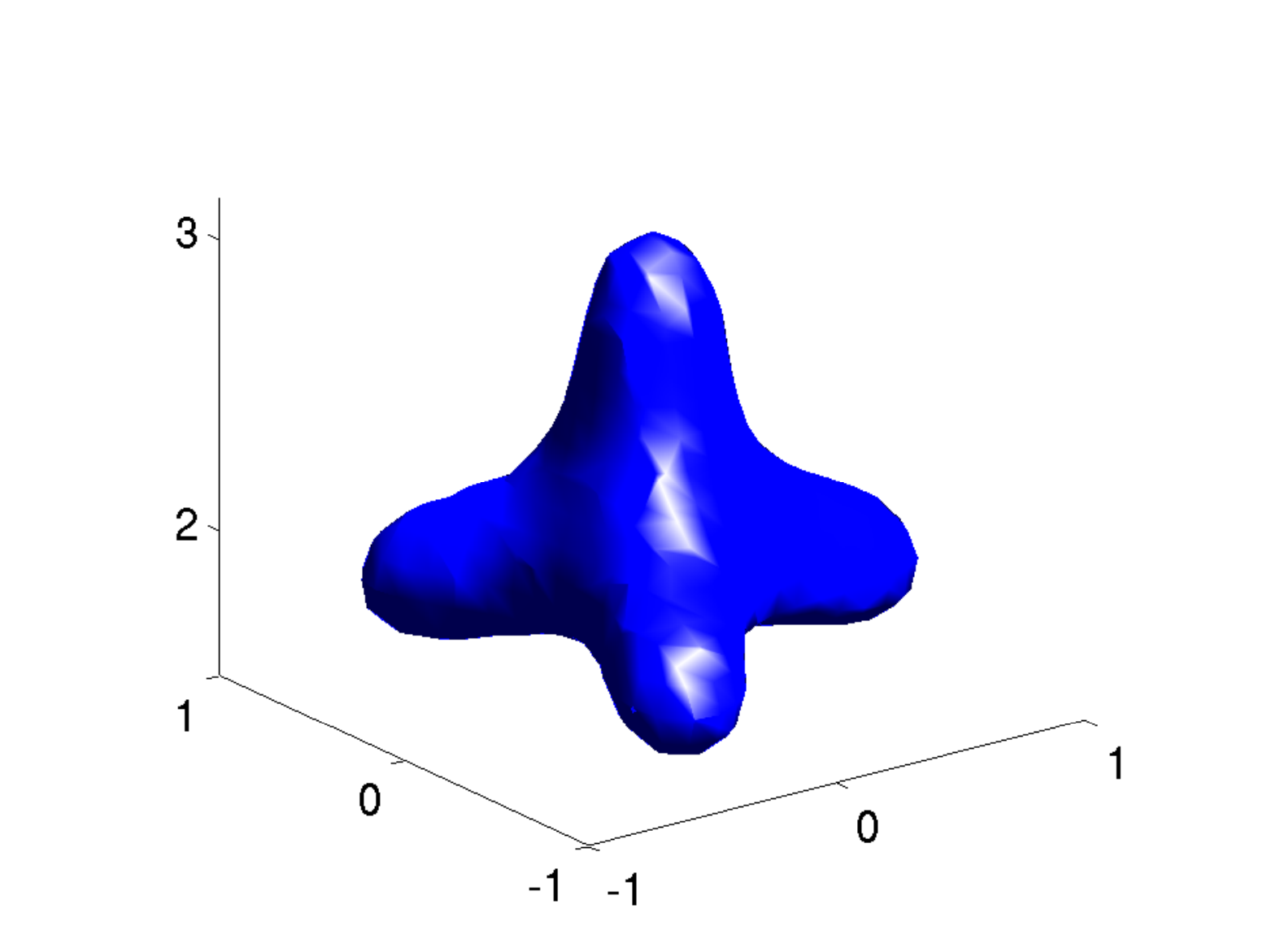}
\caption{A contour of $\text{tr}(\hat{\Phi}^2)$ for the $C=1$, $W=2+\sqrt{3}$ ($K=4$) charge $3$ solution of 
type $\ell=1$.  This shows the Higgs field is close to zero at the centre of the tetrahedron, although the 
energy density is not peaked there (fig.~\ref{charge3energy}, but note the change of 
scale).}\label{tetrahedralzeroes}
\end{figure}

As a further similarity with monopole scattering in $\mathbb{R}^3$, we observe the appearance of an 
additional zero (termed an `antizero' in \cite{Sut97}) during the $\ell=0$ scattering process with 
$W\in\mathbb{R}$.  The motion of Higgs zeroes can thus be described as follows: three zeroes move radially 
inwards on the vertices of an equilateral triangle, falling slightly below the plane $z=\beta/3$ as they 
approach.  At some ($C$-dependent) value of $W$, a zero appears on the $z$ axis, slightly above $\beta/3$ 
(fig.~\ref{zeroantizero}).  Reducing $W$ further, the zero splits into two, moving in the positive and 
negative $z$ directions, fig.~\ref{tetrahedralzeroes}.  At some value of $W$ the downward-going zero (the 
antizero) meets the three original zeroes, resulting in the toroidal two-monopole cluster of 
fig.~\ref{charge3energy}.  However, the precise value of $W$ at which this occurs is hard to resolve 
numerically.  Details of the effect of varying $C$ on the monopole structure will be presented in \cite{Mal}.
\newpage
\section{Multiplying chains}\label{doubledchains}
In this section we investigate how the Nahm data of a monopole chain can be constructed from that of a lower 
charge chain.  This is possible when a chain can be described as a lower charge chain with a rescaled size 
$C$ and period $\beta$.  We will firstly consider a generalisation of the large $N$ limit of the 
Ercolani-Sinha solution \cite{ES89} given in \cite{HW09}.  Next, we will look at how charge $2k$ Nahm data 
with $\ell=k/2$ can be expressed as charge $k$ Nahm data with $\ell=0$, and suggest an interpretation.
\subsection{Rescaling a charge 1 chain}\label{rescaling}
Harland \& Ward \cite{HW09} considered a rescaling of the Nahm data relevant to a finite chain of $N$ 
monopoles in the limit $N\to\infty$.  In this limit, the Nahm data become infinite dimensional and operate on 
a $k$ dimensional vector of functions.  The $k\times k$ matrix corresponding to this action is the Nahm data 
of a periodic monopole.  This procedure allowed the authors to reproduce the Nahm data of monopole chains of 
charge $1$, and for the special charge $2$ configuration consisting of a charge $1$ chain of halved period.  
The resulting Nahm data is equivalent to that for $W=\text{i}$ on the submanifold $\Sigma_2^1$.  For higher 
charges, this procedure does not give a point on the surface $\Sigma_k$.  For instance, in the charge $3$ 
case we have
\begin{equation}
\Phi\,=\,\begin{pmatrix}0&\text{e}^{-\beta r/3}&\text{e}^{\beta(r/3+\text{i}t)}\\\text{e}^{\beta r/3}&0&\text{e}^{-\beta r/3}\\\text{e}^{-\beta(r/3+\text{i}t)}&\text{e}^{\beta r/3}&0\end{pmatrix}\qquad\qquad A_{\bar{s}}\,=\,\frac{\beta}{6}\begin{pmatrix}1&0&0\\0&0&0\\0&0&-1\end{pmatrix}.\label{charge3nahmdata}
\end{equation}
This solution is of interest as the only currently known explicit solution with spectral curve coefficient 
$b_1\neq0$ (see section \ref{charge3symmetries}).  In fact, the characteristic polynomial of $\Phi$ is 
$\zeta^3-3\zeta-(w+w^{-1})=0$.  This is simply the $k=3$ version of the spectral curve 
$\text{det}(w-V_1(\zeta)^k)=0$, where the holonomy of the charge $1$ chain, $V_1(\zeta)$, is taken over $k$ 
periods and satisfies $\text{tr}(V_1(\zeta))=\zeta$ and $\text{det}(V_1(\zeta))=1$.  Note that $F=0$, as 
expected for a charge $1$ monopole chain (for which the Nahm data is of rank $1$).
%
%
%
\subsection{Embedding Nahm data}
Another approach to construct higher charge chains is by embedding lower charge Nahm data along the diagonal 
of a higher rank matrix, with rescaled periods and a phase shift to ensure the resulting characteristic 
polynomial of $\Phi$ is a valid spectral curve.  This construction will in general yield Nahm data of the 
wrong periodicity, although it can readily be cast into the standard form of section \ref{solutions} by a 
change of gauge.
\subsubsection{Charge $k$ from charge 1}
The charge $1$ Nahm data is simply 
$\Phi^{(1)}=C\cosh(\beta s)$, $A^{(1)}=0$.  We form a traceless rank $2$ Hitchin Higgs field by a phase shift 
of $-1$, $\Phi'=C\cosh(\beta s/2)\sigma_3$.  We should not be worried about the anti-periodicity of $\Phi'$ if 
we notice that it is periodic with period $4\pi/\beta$, while the embedded charge $1$ monopole has the dual 
period, $\beta/2$.  Now we perform a non-periodic gauge transformation with
\begin{equation*}
g\,=\,\frac{1}{\sqrt{2}}\begin{pmatrix}1&\text{e}^{\text{i}\beta t/2}\\\text{e}^{-\text{i}\beta t/2}&-1\end{pmatrix}
\end{equation*}
resulting in
\begin{equation*}
\Phi^{(2)}\,=\,g^{-1}\Phi'g\,=\,C\cosh(\beta s/2)\begin{pmatrix}0&\text{e}^{\text{i}\beta t/2}\\\text{e}^{-\text{i}\beta t/2}&0\end{pmatrix}
\end{equation*}
which is (up to a rescaling of $C$) the appropriate Hitchin Higgs field of a charge $2$ chain, as can be 
obtained using the method of section \ref{rescaling}.  The gauge potential in the usual gauge is expected to 
be $A^{(2)}_{\bar{s}}=\beta\sigma_3/8$.  Applying the inverse gauge transformation, we find that 
$A^{(2)}_{\bar{s}}=g^{-1}A'_{\bar{s}}g+g^{-1}\partial_{\bar{s}}g$ with $A'_{\bar{s}}=A^{(2)}_{\bar{s}}$.  The 
structure of the inverse Nahm operator \eqref{invnahm} relating the symmetries of $\zeta$ and $z$ to those of 
$\Phi$ and $A$ allows us to interpret the embedded charge $1$ Nahm data as describing two monopoles of the 
same orientation (due to the rotational symmetry $(\zeta,z)\sim(-\zeta,z)$) but with $z$ positions shifted by 
$\pm\beta/4$ from the origin (this is determined from \eqref{invnahm} as twice the shift in 
$A^{(1)}_{\bar{s}}$ from $A^{(1)}_{\bar{s}}=0$ for the single chain centered at $z=0$).
\par An analogous procedure can be carried out to construct the charge $3$ chain of section \ref{rescaling} 
from charge $1$ Nahm data.  This time we have
\begin{equation*}
\Phi'\,=\,2\,\text{diag}\left(\cosh\left(\frac{\beta s}{3}\right),\cosh\left(\frac{\beta s+2\text{i}\pi}{3}\right),\cosh\left(\frac{\beta s-2\text{i}\pi}{3}\right)\right),\,A_{\bar{s}}\,=\,\frac{\beta}{6}\text{diag}(1,0,-1)
\end{equation*}
which is gauge equivalent to \eqref{charge3nahmdata} by conjugation with
\begin{equation*}
g\,=\,\frac{1}{\sqrt{3}}\begin{pmatrix}1&\text{e}^{\text{i}\beta t/3}&\text{e}^{2\text{i}\beta t/3}\\\text{e}^{-\text{i}\beta t/3-2\text{i}\pi/3}&1&\text{e}^{\text{i}\beta t/3+2\text{i}\pi/3}\\\text{e}^{-2\text{i}\beta t/3-2\text{i}\pi/3}&\text{e}^{-\text{i}\beta t/3+2\text{i}\pi/3}&1\end{pmatrix}.
\end{equation*}
\subsubsection{Charge 4 from charge 2}
The same idea can be applied to higher charges.  This allows us to take, say, a charge $2$ monopole in pairs 
to give charge $4$ Nahm data where the Higgs field is block-diagonal,
\begin{equation*}
\Phi^{(4)}\,=\,\begin{pmatrix}\Phi^{(2)}&0\\0&\Phi'^{(2)}\end{pmatrix},
\end{equation*}
which has a valid spectral curve as long as both $\Phi^{(2)}$ and $\Phi'^{(2)}$ have the same $\ell$, with a 
relative overall phase of $\text{e}^{\text{i}\pi/2}$ and with $K$ of opposite signs.
\par A special case is provided by $\Phi^{(2)}$ with $\ell=0$ and $K=0$.  The gauge transformation
\begin{equation*}
g\,=\,\frac{1}{\sqrt{2}}\begin{pmatrix}1&0&\text{e}^{\text{i}\beta t/2}&0\\0&1&0&\text{e}^{\text{i}\beta t/2}\\\text{e}^{-\text{i}\beta t/2}&0&-1&0\\0&\text{i}\text{e}^{-\text{i}\beta t/2}&0&-\text{i}\end{pmatrix}
\end{equation*}
shows that this is equivalent to the charge $4$ case with $\ell=2$ and $W=1$ (see section 
\ref{symmetricsplitting}).  In other words, there are particular charge $4$ configurations which can be 
understood as charge $2$ chains ``in disguise'', a result which could be anticipated by the $\mathbb{Z}_4$ 
symmetry of both cases.
\par The decoupling of the Nahm data into block-diagonal form suggests the relevant monopoles are `maximally 
separated' and non-interacting.  
%
%
This is reminiscent of the decoupling of the moduli space metric of a charge $2$ monopole into a direct 
product of two $1$-monopole moduli space 
metrics for two well separated monopoles, \cite{GM95,Bie08}.
\section{Conclusions}
The work presented in this paper extends the construction of periodic monopoles of charge $1$ and $2$ to a 
particular family of higher charges.  The resulting scattering processes correspond (in the small size to 
period ratio) to those scattering processes in $\mathbb{R}^3$ with cyclic symmetry \cite{Sut97}, and it is 
clear how the number of different possibilites of a charge $k$ monopole splitting into two clusters arises 
naturally from the structure of the Hitchin fields.  We also conisder a special case in which equally charged 
clusters are maximally separated, allowing the Nahm data to decouple and a description to be made in terms of 
lower charge monopoles.
\par There remain further geodesic submanifolds that do not fit the Ansatz \eqref{Phin}.  In particular, the 
search is still underway for solutions of the Hitchin equations with $b_1\neq0$ and $b_0=0$ 
(fig.~\ref{geo3}), for which only the example \eqref{charge3nahmdata} has been identified.  A related 
question is whether one can construct a charge $k$ `twisted chain' of equally spaced monopoles invariant 
under a joint rotation and shift: $(\zeta,z)\sim(\text{e}^{\text{i}\pi/k}\zeta,z+m\beta/k)$ for $0\leq m<k$.  
The existence of monopole chains with this symmetry has been established \cite{Har}, but explicit periodic 
solutions are only available for $m=0$ and $m=k/2$: they are the points on $\Sigma_k^m$ with $K=0$ 
($W=\pm\text{i}$).  For $m=1$, $k=3$ the spectral curve is expected to have $b_1=b_0=0$ (central panel of 
fig.~\ref{geo3}).  It would be interesting to consider whether this solution can be described as a point on 
the geodesic containing the `tripled chain' of section \ref{rescaling} and fig.~\ref{geo3}.
\section{Acknowledgements}
Thanks to D.G.~Harland for discussions and to R.S.~Ward for use of the numerical procedure of \cite{HW09}.  
This work was funded by an STFC studentship.


\newpage
{\small\begin{thebibliography}{99}
\raggedright
\bibitem{ChK01}S.~Cherkis, A.~Kapustin, \emph{Nahm Transform for Periodic Monopoles and N=2 Super Yang-Mills Theory}, Commun.~Math.~Phys.~{\bf 218} (2001) 333, \href{http://arxiv.org/abs/hep-th/0006050}{\texttt{arXiv:hep-th/0006050}}
\bibitem{ChK03}S.A.~Cherkis, A.~Kapustin, \emph{Periodic Monopoles With Singularities And N=2 Super-QCD}, Commun.~Math.~Phys.~{\bf 234} (2003) 1, \href{http://arxiv.org/abs/hep-th/0011081}{\texttt{arXiv:hep-th/0011081}}
\bibitem{HW09}D.~Harland, R.S.~Ward, \emph{Dynamics of periodic monopoles}, Phys.~Lett.~{\bf B 675} (2009) 262, \href{http://arxiv.org/abs/0901.4428}{\texttt{arXiv:0901.4428 [hep-th]}}
\bibitem{War05}R.S.~Ward, \emph{Periodic monopoles}, Phys.~Lett.~{\bf B 619} (2005) 177, \href{http://arxiv.org/abs/hep-th/0505254}{\texttt{arXiv:hep-th/0505254}}
\bibitem{Mal13}R.~Maldonado, \emph{Periodic monopoles from spectral curves}, JHEP02(2013)099, \href{http://arxiv.org/abs/1212.4481}{\texttt{arXiv:1212.4481 [hep-th]}}
\bibitem{HMM95}N.J.~Hitchin, N.S.~Manton, M.K.~Murray, \emph{Symmetric monopoles}, Nonlinearity {\bf 8} (1995) 661, \href{http://arxiv.org/abs/dg-ga/9503016}{\texttt{arXiv:dg-ga/9503016}}
\bibitem{Sut97}P.M.~Sutcliffe, \emph{Cyclic monopoles}, Nucl.~Phys.~{\bf B 505} (1997) 517, \href{http://arxiv.org/abs/hep-th/9610030}{\texttt{arXiv:hep-th/9610030}}
\bibitem{MW}R.~Maldonado, R.S.~Ward, \emph{Geometry  of Periodic Monopoles}, to appear in Phys.~Rev.~{\bf D}, \href{http://arxiv.org/abs/1309.7013}{\texttt{arXiv:1309.7013 [hep-th]}}
\bibitem{Sut96}P.M.~Sutcliffe, \emph{Seiberg-Witten theory, monopole spectal curves and affine Toda solitons}, Phys.~Lett.~{\bf B 381} (1996) 129, \href{http://arxiv.org/abs/hep-th/9605192}{\texttt{arXiv:hep-th/9605192}}
\bibitem{Bra11}H.W. Braden, \emph{Cyclic Monopoles, Affine Toda and Spectral Curves}, Commun. Math. Phys. {\bf 308} (2011) 303, \href{http://arxiv.org/abs/1002.1216}{\texttt{arXiv:1002.1216 [math-ph]}}
\bibitem{Mal}R.~Maldonado, \emph{Scaling limits of periodic monopoles}, in preparation
\bibitem{ES89}N.~Ercolani, A.~Sinha, \emph{Monopoles and Baker Functions}, Commun.~Math.~Phys.~{\bf 125} (1989) 385
\bibitem{GM95}G.W.~Gibbons, N.S.~Manton, \emph{The moduli space metric for well-separated BPS monopoles}, Phys.~Lett.~{\bf B 356} (1995) 32
\bibitem{Bie08}R.~Bielawski, \emph{Monopoles and Clusters}, Commun.~Math.~Phys.~{\bf 284} (2008) 675, \href{http://arxiv.org/abs/hep-th/0702190}{\texttt{arXiv:hep-th/0702190}}
\bibitem{Har}D.G.~Harland, \emph{Cyclic Monopole Chains}, to appear
\end{thebibliography}}
\end{document}